\documentclass[aoas,MSNbibl,nameyear,dvips]{arximspdf}
\usepackage{dcolumn}
\usepackage{graphicx}

\doi{10.1214/14-AOAS762} 
\volume{8}
\issue{3}
\pubyear{2014}
\firstpage{1395}
\lastpage{1415}
\docsubty{FLA}

\makeatletter
\newcolumntype{d}[1]{D{.}{.}{#1}}
\makeatother

\begin{document}
\begin{frontmatter}

\title{Inference for deformation and interference in~3D~printing}
\runtitle{Deformation and interference in 3D printing}

\begin{aug}
\author[A]{\fnms{Arman}~\snm{Sabbaghi}\corref{}\ead[label=e1]{armansabbaghi.stat@gmail.com}\thanksref{T1,M1}},
\author[B]{\fnms{Tirthankar}~\snm{Dasgupta}\ead[label=e2]{dasgupta@stat.harvard.edu}\thanksref{T2,M2}},
\author[C]{\fnms{Qiang}~\snm{Huang}\ead[label=e3]{qiang.huang@usc.edu}\thanksref{T3,M3}}
\and
\author[C]{\fnms{Jizhe}~\snm{Zhang}\ead[label=e4]{jizhezha@usc.edu}\thanksref{M3}}
\runauthor{Sabbaghi, Dasgupta, Huang and Zhang}
\affiliation{Purdue University\thanksmark{M1}, Harvard University\thanksmark{M2} and
University of Southern California\thanksmark{M3}}
\address[A]{A. Sabbaghi \\
Department of Statistics \\
Purdue University\\
250 N. University Street \\
West Lafayette, Indiana 47907 \\
USA\\
\printead{e1}} 
\address[B]{T. Dasgupta \\
Department of Statistics \\
Harvard University\\
1 Oxford Street, 7th Fl. \\
Cambridge, Massachusetts 02138 \\
USA\\
\printead{e2}}
\address[C]{Q. Huang \\
J. Zhang \\
Daniel J. Epstein Department of Industrial\\
\quad and Systems Engineering \\
University of Southern California\\
3715 McClintock Avenue \\
Los Angeles, California 90089 \\
USA\\
\printead{e3} \\
\phantom{\textsc{E-mail}: }\printead*{e4}}
\end{aug}

\thankstext{T1}{Supported by National Science Foundation Grant DGE-1144152.}
\thankstext{T2}{Supported by National Science Foundation Grant DMS-11-07004.}
\thankstext{T3}{Supported by Office of Naval Research Grant \#N000141110671.}

\received{\smonth{11} \syear{2013}}
\revised{\smonth{4} \syear{2014}}

%
\begin{abstract}
Additive manufacturing, or 3D printing, is a promising manufacturing
technique marred by product deformation due to material solidification
in the printing process. Control of printed product deformation can be
achieved by a compensation plan. However, little attention has been
paid to interference in compensation, which is thought to result from
the inevitable discretization of a compensation plan. We investigate
interference with an experiment involving the application of
discretized compensation plans to cylinders. Our treatment illustrates
a principled framework for detecting and modeling interference, and
ultimately provides a new step toward better understanding quality
control for 3D printing.
\end{abstract}

%
\end{frontmatter}

\section{Interference in compensation}
\label{sec1}

Additive manufacturing, or 3D printing, refers to a class of technology
for the direct fabrication of physical products from 3D Computer-Aided
Design (CAD) models. In contrast to material removal processes in
traditional machining, the printing process adds material layer by
layer. This enables direct printing of geometrically complex products
without affecting building efficiency. No extra effort is necessary for
molding construction or fixture tooling design, making 3D printing a
promising manufacturing technique [\citet{hilton2000rapid,gibson2009additive,melchels2010review,campbell2011could}].
Despite
these promising features, accurate control of a product's printed
dimensions remains a major bottleneck. Material solidification during
layer formation leads to product deformation, or shrinkage
[\citet{wang1996influence}], which reduces the utility of printed
products. Shrinkage control is crucial to overcome the accuracy barrier
in 3D printing.

To control detailed features along the boundary of a printed product,
\citet{tong2003parametric} and \citet{tong2008error} used polynomial
regression models to first analyze shrinkage in different directions
separately, and then compensate for product deformation by changing the
original CAD accordingly. Unfortunately, their predictions are
independent of the product's geometry, which is not consistent with the
physical manufacturing process. \citet{huangzhangsabbaghidasgupta}
built on this work, establishing a generic, physically consistent
approach to model and predict product deformations, and to derive
compensation plans. The essence of this new modeling approach is to
transform in-plane geometric errors from the Cartesian coordinate
system into a functional profile defined on the polar coordinate
system. This representation decouples the geometric shape complexity
from the deformation modeling, and a generic formulation of shape
deformation can thus be achieved. The approach was developed for a
stereolithography process, and in experiments achieved an improvement
of one order of magnitude in reduction of deformation for cylinder
products.

\begin{figure}

\includegraphics{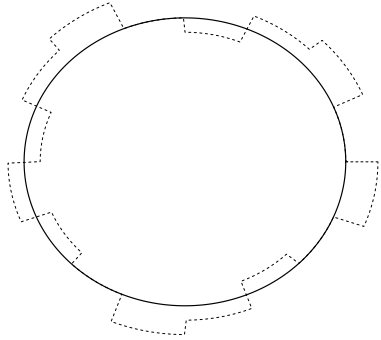}

\caption{A discretized compensation plan (dashed line) to the nominal
boundary (solid line). Note that compensation could be negative.}
\label{compensationexample}
\end{figure}

However, an important issue not yet addressed in the previously cited
work on deformation control for 3D printing is how the application of
compensation to one section of a product will affect the deformation of
its neighbors. Compensation plans are always discretized according to
the tolerance of the 3D printer, in the sense that sections of the CAD
are altered by single amounts, for example, as in Figure~\ref{compensationexample}. Furthermore, when planning an experiment to
assess the effect of compensation on product deformation, it is natural
to discretize the quantitative ``compensation'' factor into a finite
number of levels, which also leads to a product having a more complex
boundary. Ultimately, such changes may introduce interference between
different sections of the printed product, which is defined to occur
when one section's deformation depends not only on its assigned
compensation, but also on compensations assigned to its neighbors
[\citet{Rubin1980}]. For example, in Figure~\ref{compensationexample},
the deformation for points near the boundary of two neighboring
sections should depend on compensations applied to both. By the same
logic, interference becomes a practical issue when printing products
with complex geometry. Therefore, to improve quality control in 3D
printing, it is important to formally investigate complications
introduced by the interference that results from discretization in
compensation plans. We take the first step with an experiment involving
a discretized compensation plan for a simple shape.

We begin in Section~\ref{sec2} with a review of interference, models
for product
deformation, and the effect of compensation. Adoption of the Rubin
Causal Model
[RCM, \citet{holland1986}] is a significant and novel feature of our
investigation,
and facilitates the study of interference. Section~\ref{secnocompensationfit}
summarizes the basic model and analysis for deformation of cylinders
given by
\citet{huangzhangsabbaghidasgupta}. Our analyses are in Sections~\ref{secexperimentaldesign}--\ref{secrefinedmodelinterference}: we first
describe an experiment hypothesized to generate interference, then
proceed with
posterior predictive checks to demonstrate the existence of
interference, and finally
conclude with a model that captures interference. A statistically
substantial idea in
Section~\ref{secassessinginterference} is that, in experiments with
distinct units of
analysis and units of interpretation  [\citet{CoxDonnelly}, pages
18--19], the
posterior distribution of model parameters, based on ``benchmark''
data, yields a
simple assessment and inference for interference in the experiment,
similar to that
suggested by \citet{sobel} and \citet{rosenbaum}. Analyses in Sections~\ref{secsimplemodelinterference}--\ref{secrefinedmodelinterference}
demonstrate how discretized compensation plans complicate quality
control through the
\hyperref[sec1]{Introduction} of interference. This illustrates the fact that in complex
manufacturing
processes, a proper definition of experimental units and understanding
of interference
are critical to quality control.

\section{Potential outcomes and interference}
\label{sec2}

\subsection{Experimental units and potential outcomes}
\label{secunitsoutcomes}

We use the general\break framework for product deformation given by
\citeauthor{huangzhangsabbaghidasgupta} [(\citeyear{huangzhangsabbaghidasgupta}), pages 3--6].
Suppose a product
has intended shape $\psi_0$ and observed shape $\psi$ under a 3D
printing process. Deformation is informally described as the difference
between $\psi$ and $\psi_0$, where we can represent both either in the
Cartesian coordinate system $(x,y,z)$ or cylindrical coordinate system
$(r,\theta,z)$. Cylindrical coordinates facilitate deformation modeling
and are used throughout.

For illustrative purposes, we define terms for two-dimensional products
(notation for three dimensions follows immediately). Quality control
requires an understanding of deformation in different regions of the
product that receive different amounts of compensation. We therefore
define a finite number $N$ of points on the boundary of the product,
corresponding to specific angles $\theta_1, \ldots, \theta_N$, as the
experimental units. The desired boundary from the CAD model is defined
by the function $r_0(\theta)$, denoting the nominal radius at angle
$\theta$. We consider only one (quantitative) treatment factor,
compensation to the CAD, defined as a change in the nominal radius of
the CAD by $x_i$ units at $\theta_i$ for $i = 1, \ldots, N$.
Compensation is not restricted to be nonnegative. The potential radius
for $\theta_i$ under compensation $\mathbf{x} = (x_1, \ldots, x_N)$ to
$\theta_1, \ldots, \theta_N$ is a function of $\theta_i$, $r_0(\cdot)$,
and $\mathbf{x}$, denoted by $r(\theta_i, r_0(\cdot), \mathbf{x})$. The
difference between the potential and nominal radius at $\theta_i$
defines deformation, and so
%
\begin{equation}
\label{eqpotentialoutcomes}
\Delta r\bigl(\theta_i, r_0(\cdot),
\mathbf{x}\bigr) = r\bigl(\theta_i, r_0(\cdot),
\mathbf{x}\bigr) - r_0(\theta_i)
\end{equation}
is defined as our potential outcome for $\theta_i$. Potential outcomes
are viewed as fixed numbers, with randomness introduced in Section~\ref{secmodeling} in our general model for the potential outcomes.

\begin{figure}

\includegraphics{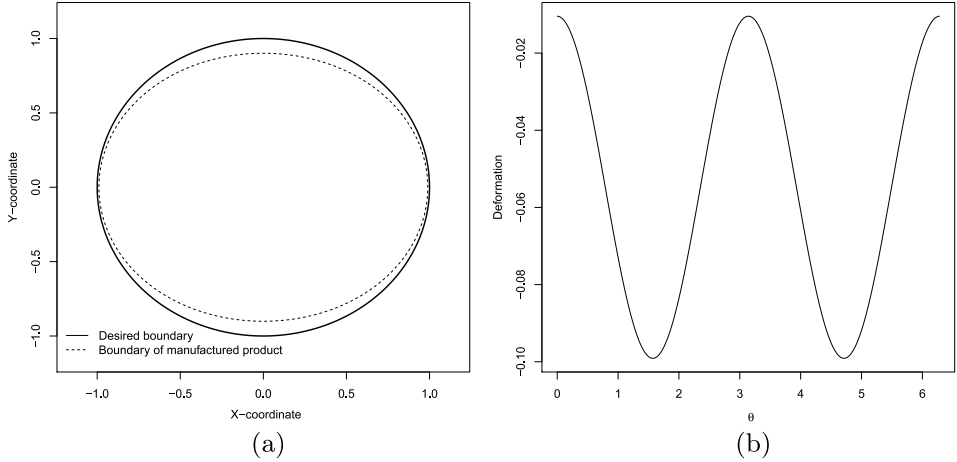}

\caption{(\textup{a}) Ideal shape (solid line) versus the actual shape
(dashed line). \textup{(b)} Visualization of shrinkage.}
\label{deformationcurve}
\end{figure}

This definition of the potential outcome is convenient for visualizing
shrinkage. For example, suppose the desired shape of the product is the
solid line, and the manufactured product when $\mathbf{x} = \mathbf{0}
= (0, \ldots, 0)$ is the dashed line, in Figure~\ref{deformationcurve}(a). Plotting the deformation at each angle yields a
visualization amenable to analysis [Figure~\ref{deformationcurve}(b)].
Orientation is fixed: we match the coordinate axes of the printed
product with those of the CAD model.

\subsection{Interference}
\label{secinterference}

A unit $\theta_i$ is said to be affected by interference if
\[
\Delta r\bigl(\theta_i, r_0(\cdot), \mathbf{x}\bigr)
\neq\Delta r\bigl(\theta_i, r_0(\cdot),
\mathbf{x}'\bigr)
\]
for at least one pair of distinct treatment vectors $\mathbf{x}, \mathbf
{x}' \in\mathbb{R}^N$ with $x_i = x'_i$
[\citet{Rubin1980}].
If there is no interference, then $\Delta r(\theta
_i, r_0(\cdot), \mathbf{x})$ is a function of $\mathbf{x}$ only via the
component $x_i$. As the experimental units reside on a connected
boundary, the deformation of one unit may depend on compensations
assigned to its neighbors when the compensation plan is discretized.
Perhaps less plausible, but equally serious, is the possible leakage of
assigned compensations across units. These considerations explain the
presence of the vector $\mathbf{x}$, containing compensations for all
units, in the potential outcome notation (\ref{eqpotentialoutcomes}).
Practically, accommodations made for interference should reduce bias in
compensation plans for complex products and improve quality control.

\subsection{General deformation model}
\label{secmodeling}

Following \citeauthor{huangzhangsabbaghidasgupta}
[(\citeyear{huangzhangsabbaghidasgupta}),
pages 6--8],  our
potential outcome model under compensation plan $\mathbf{x} = \mathbf
{0}$ is decomposed into three components:
%
\begin{equation}
\label{eqdecomp1}
\Delta r\bigl(\theta_i, r_0(\cdot),
\mathbf{0}\bigr) = f_1\bigl(r_0(\cdot)\bigr) +
f_2\bigl(\theta_i, r_0(\cdot), \mathbf{0}
\bigr) + \varepsilon_{i}.
\end{equation}
Function $f_1(r_0(\cdot))$ represents average deformation of a given
nominal shape $r_0(\cdot)$ independent of location $\theta_i$, and
$f_2(\theta_i, r_0(\cdot), \mathbf{0})$ is the additional
location-dependent deformation, geometrically and physically related to
the CAD model. We can also interpret $f_1(\cdot)$ as a low-order
component and $f_2(\cdot, \cdot, \mathbf{0})$ as a high-order component
of deformation. The $\varepsilon_{i}$ are random variables representing
high-frequency components that add on to the main trend, with
expectation $\mathbb{E}(\varepsilon_{i}) = 0$ and $\operatorname{Var}(\varepsilon
_{i}) < \infty$ for all $i = 1, \ldots, N$.

Figure~\ref{deformationcurve} demonstrates model (\ref{eqdecomp1}).
In this example, $r_0(\theta) = r_0$, so $f_1(\cdot)$ is a function of
$r_0$, and $f_2(0, r_0, \mathbf{0}) = f_2(2\pi, r_0, \mathbf{0})$.
Decomposition of deformation into lower and higher order terms yields
%
\begin{equation}
\label{eqfourierseries} \Delta r(\theta_i, r_0, \mathbf{0}) =
c_{r_0} + \sum_k \bigl\{
a_{r_0, k} \cos(k\theta_i) + b_{r_0, k} \sin (k
\theta_i) \bigr\} + \varepsilon_{i},
\end{equation}
where $f_1(r_0) = c_{r_0}$, and $\{a_{r_0, k},  b_{r_0, k}\}$ are
coefficients of a Fourier series expansion of $f_2(\cdot, \cdot, \mathbf
{0})$. The $\{ a_{r_0, k},  b_{r_0, k} \}$ terms with large $k$
represent the product's surface roughness, which is not of primary interest.

\subsection{General compensation and interference models}
\label{seccompensation}

Under the polar coordinate system, a compensation of $x_i$ units at
$\theta_i$ can be thought of as an extension of the product's radius by
$x_i$ units in that direction. Bearing this in mind, we first
follow   \citeauthor{huangzhangsabbaghidasgupta} [(\citeyear{huangzhangsabbaghidasgupta}), page 8] to extend (\ref{eqdecomp1})
to accommodate compensations, and then build upon this to
give an extension that can help capture interference resulting from
discretized compensation plans.

Let $r(\theta_i, r_0(\cdot), (x_i, \ldots, x_i)) = r(\theta_i, r_0(\cdot
), x_i\mathbf{1})$ denote the potential radius for $\theta_i$ under
compensation of $x_i$ units to all points. Compensation $x_i\mathbf{1}$
is equivalent, in terms of the final manufactured product, as if a CAD
model with nominal radius $r_0(\cdot) + x_i$ and compensation $\mathbf{0}$ was initially submitted to the 3D printer. Then
%
\begin{eqnarray}
\label{eqintstep1}
r\bigl(\theta_i, r_0(\cdot),
x_i\mathbf{1}\bigr) - \bigl\{ r_0(\theta_i)
+ x_i \bigr\} &=& r\bigl(\theta_i, r_0(
\cdot) + x_i, \mathbf{0}\bigr) - \bigl\{ r_0(\theta
_i) + x_i \bigr\}
\nonumber
\\[-8pt]
\\[-8pt]
\nonumber
&=& \Delta r\bigl(\theta_i, r_0(\cdot) +
x_i, \mathbf{0}\bigr),
\end{eqnarray}
where $\Delta r(\theta_i, r_0(\cdot) + x_i, \mathbf{0})$ follows the
same form as (\ref{eqdecomp1}), abbreviated as
%
\begin{equation}
\label{eqintstep2} \Delta r\bigl(\theta_i, r_0(\cdot) +
x_i, \mathbf{0}\bigr) = \mathbb{E} \bigl\{ \Delta r\bigl(
\theta_i, r_0(\cdot) + x_i, \mathbf{0}
\bigr) \bigr\} + \varepsilon_i.
\end{equation}
Consequently, the potential outcome for $\theta_i$ is
%
\begin{eqnarray}
\label{eqmodelcomp2}
\Delta r\bigl(\theta_i, r_0(\cdot),
x_i\mathbf{1}\bigr) &=& r\bigl(\theta_i,
r_0(\cdot ), x_i\mathbf{1}\bigr) - r_0(
\theta_i)
\nonumber
\\
&=& r\bigl(\theta_i, r_0(\cdot), x_i
\mathbf{1}\bigr) - \bigl\{ r_0(\theta_i) +
x_i \bigr\} + x_i
\nonumber
\\[-8pt]
\\[-8pt]
&=& \Delta r\bigl(\theta_i, r_0(\cdot) +
x_i, \mathbf{0}\bigr) + x_i
\nonumber
\\
&=& \mathbb{E} \bigl\{ \Delta r\bigl(\theta_i, r_0(
\cdot) + x_i, \mathbf{0}\bigr) \bigr\} + x_i +
\varepsilon_i.\nonumber
\end{eqnarray}
The last two steps follow from (\ref{eqintstep1}) and (\ref
{eqintstep2}), respectively. If $x_i$ is small relative to
$r_0(\theta_i)$, then (\ref{eqmodelcomp2}) can be approximated using
the first and second terms of the Taylor expansion of
$\mathbb{E}  \{ \Delta r(\theta_i, r_0(\cdot) + x_i, \mathbf{0})
 \}$ at $r_0(\theta_i)$:
%
\begin{eqnarray}
\label{eqcomptaylor}
\hspace*{2pt}\quad\Delta r\bigl(\theta_i, r_0(\cdot),
x_i\mathbf{1}\bigr) &\approx & \mathbb{E} \bigl\{ \Delta r\bigl(
\theta_i, r_0(\cdot), \mathbf{0}\bigr) \bigr\}
\nonumber
\\
&&{}+ (x_i - 0) \biggl[ \frac{d}{dx} \mathbb{E} \bigl\{ \Delta
r\bigl(\theta_i, r_0(\cdot) + x, \mathbf{0}\bigr) \bigr
\} \biggr]_{x = 0} + x_i + \varepsilon _{i}
\\
&=& \Delta r\bigl(\theta_i,r_0(\cdot),\mathbf{0}\bigr)
+ \bigl\{ 1 + h\bigl(\theta_i, r_0(\cdot), \mathbf{0}
\bigr)\bigr\} x_i,
\nonumber
\end{eqnarray}
where $h(\theta_i, r_0(\cdot), \mathbf{0}) =  [ d/dx  \mathbb{E}
 \{ \Delta r(\theta_i, r_0(\cdot) + x, \mathbf{0})  \}
 ]_{x = 0}$. Under a specified parametric model for the potential
outcomes, this Taylor expansion is performed conditional on the model
parameters. When there is no interference,
\[
\Delta r\bigl(\theta_i, r_0(\cdot), \mathbf{x}\bigr) =
\Delta r\bigl(\theta_i, r_0(\cdot), x_i
\mathbf{1}\bigr)
\]
for any $\mathbf{x} \in\mathbb{R}^N$, and so (\ref{eqcomptaylor}) is
a model for compensation effects in this case.

We can generalize this model to incorporate interference in a simple
manner for a compensation plan $\mathbf{x}$ with different units
assigned different compensations. As all units are connected on the
boundary of the product, unit $\theta_i$'s treatment effect will change
due to interference from its neighbors, so that $\theta_i$ will deform
not just according to its assigned compensation $x_i$, but instead
according to a compensation $g_i(\mathbf{x})$. Thus, we generalize (\ref
{eqcomptaylor}) to
%
\begin{equation}
\label{eqcomptaylorinterference}
\Delta r\bigl(\theta_i, r_0(\cdot),
\mathbf{x}\bigr) \approx \Delta r\bigl(\theta_i ,r_0(
\cdot), \mathbf{0}\bigr) + \bigl\{ 1 + h\bigl(\theta_i,
r_0(\cdot), \mathbf{0}\bigr) \bigr\} g_i(\mathbf{x}),
\end{equation}
where the \textit{effective treatment} $g_i(\mathbf{x})$ is a function of
$x_i$ and assigned compensations for neighbors of $\theta_i$ (with the
definition of neighboring units naturally dependent on the specific
product), hence potentially a function of the entire vector $\mathbf
{x}$. Allowing the treatment effect for $\theta_i$ to depend on
treatments assigned to its neighboring units effectively incorporates
interference in a meaningful manner, as will be seen in the analysis of
our experiment.

\section{Experimental design and analysis for interference}
\label{sec3}

\subsection{Compensation model for cylinders}
\label{secnocompensationfit}

Huang et~al. [(\citeyear{huangzhangsabbaghidasgupta}), page 12]   constructed four
cylinders with $r_0 = 0.5, 1, 2$, and $3$ inches, and used $N_{0.5} =
749, N_1 = 707, N_2 = 700$, and $N_3 = 721$ equally-spaced units from
each. Based on the logic in Section~\ref{secmodeling}, they fitted
%
\begin{equation}\label{eqnocompmodel}
\Delta r(\theta_i, r_0, \mathbf{0}) = x_0
+ \alpha(r_0+x_0)^a + \beta(r_0+x_0)^b
\cos(2 \theta_i) + \varepsilon_{i}
\end{equation}
to the data, with $\varepsilon_{i} \sim\mathrm{N}(0, \sigma^2)$
independently, and parameters $\alpha, \beta, a, b, x_0$, and $\sigma$
independent of $r_0$. Specifically, for the cylinder, the
location-independent term is thought to be proportional to $r_0$, so
that with overexposure of $x_0$ units it would be of the form $x_0 +
\alpha(r_0 + x_0)$. Furthermore, the location-dependent term is thought
to be a harmonic function of $\theta_i$, and also proportional to
$r_0$, of the form $\beta(r_0 + x_0)\cos(2\theta_i)$ with
overexposure. Independent errors are used throughout because the focus
is on a correct specification of the mean trend in deformation
(Appendix~\ref{seccorrelation} contains a discussion on this point).
\citet{huangzhangsabbaghidasgupta} specified
\[
a \sim\mathrm{N}\bigl(1, 2^2\bigr), \qquad b \sim\mathrm{N}
\bigl(1,1^2\bigr), \qquad \log (x_0) \sim\mathrm{N}
\bigl(0,1^2\bigr)
\]
and placed flat priors on $\alpha, \beta$, and $\log (\sigma)$,
with all parameters independent {a priori}. Posterior draws of
the parameters were obtained by Hamiltonian Monte Carlo [HMC,
\citet{duanehybrid1987}] and are summarized in Table~\ref{posteriorpredictivetable}, with convergence diagnostics discussed in
Appendix~\ref{secdiagnostics}. A simple comparison of the posterior
predictive distribution of product deformation to the observed data
[\citet{huangzhangsabbaghidasgupta}, page 19] demonstrates the good fit,
and so we proceed with this specification and parameter inferences to
design and analyze an experiment for interference.

%
%

\begin{table}
\caption{Summary of 1000 posterior draws of parameters after a
burn-in of 500  when no compensation is applied. This is drawn from
Table~5 in \citet{huangzhangsabbaghidasgupta}. Effective sample size
is abbreviated as ESS throughout}\label{posteriorpredictivetable}
\begin{tabular*}{\textwidth}{@{\extracolsep{\fill}}ld{2.9}d{1.9}d{2.9}d{3.15}c@{}}
\hline
                 & \multicolumn{1}{c}{\textbf{Mean}} & \multicolumn{1}{c}{\textbf{SD}} & \multicolumn{1}{c}{\textbf{Median}} & \multicolumn{1}{c}{$\bolds{95\%}$
                 \textbf{credible} \textbf{interval}} & \multicolumn{1}{c@{}}{\textbf{ESS}} \\\hline
$\alpha$         & -1.34 \times 10^{-2}              &  1.6 \times 10^{-4}              &   -1.34 \times 10^{-2}                & (-1.37,  -1.31) \times 10^{-2}                                     & $8198$                              \\
$\beta$          &  5.7 \times 10^{-3}                &  3.1 \times 10^{-5}             &   5.71 \times 10^{-3}                 & (5.65, 5.8) \times 10^{-3}                                        & $9522$                              \\
$a$              & 8.61 \times 10^{-1}               &  7.33 \times 10^{-3}             &   8.61 \times 10^{-1}                 & (8.47, 8.75) \times 10^{-1}                                       & $8223$                              \\
$b$              & 1.13                             &  5.46 \times 10^{-3}             &   1.13                               & (1.12, 1.14)                                                       & $9424$                              \\
$x_0$            & 8.79 \times 10^{-3}              &  1.5 \times 10^{-4}              &  8.79 \times 10^{-3}                 & (8.5, 9.07) \times 10^{-3}                                         & $8211$                              \\
$\sigma$         & 8.7 \times 10^{-4}               &  1.18 \times 10^{-5}            &  8.7 \times 10^{-4}                  & (8.5, 8.9) \times 10^{-4}                                         & $9384$                              \\\hline
\end{tabular*}
\end{table}

Substituting $\Delta r(\theta_i, r_0, \mathbf{0})$ from (\ref{eqnocompmodel}) into the general model (\ref{eqmodelcomp2}), we have
%
\begin{eqnarray}
\label{compcylinderorigin}
&& \Delta r(\theta_i, r_0,
x_i\mathbf{1})
\nonumber
\\[-8pt]
\\[-8pt]
\nonumber
&& \qquad = x_0 + x_i +
\alpha(r_0 + x_0 + x_i)^a +
\beta(r_0 + x_0 + x_i)^b \cos(2
\theta_i) + \varepsilon_{i}.
\end{eqnarray}
The Taylor expansion at $r_0 + x_0$, as in (\ref{eqcomptaylor}),
yields the model
%
\begin{eqnarray}
\label{eqmodelcompcylinder} &&\Delta r(\theta_i, r_0,
x_i\mathbf{1}) \nonumber\\
&& \qquad = x_0 + \alpha(r_0 +
x_0)^a + \beta(r_0 + x_0)^b
\cos(2 \theta_i)
\\
&&\qquad\quad {}+ \bigl\{ 1 + a\alpha(r_0 + x_0)^{a-1}
+ b\beta(r_0 + x_0)^{b-1}\cos(2
\theta_i) \bigr\}x_i + \varepsilon_{i}.\nonumber
\end{eqnarray}
We incorporate interference for a plan $\mathbf{x}$ with different
units assigned different compensations by changing $x_i$ in the right
side of (\ref{eqmodelcompcylinder}) to $g_i(\mathbf{x})$, with the
functional form of $g_i(\mathbf{x})$ derived by exploratory means in
Section~\ref{secassessinginterference}.

\subsection{Experimental design for interference}
\label{secexperimentaldesign}

Under a discretized compensation plan, the boundary of a product is
divided into sections, with all points in one section assigned the same
compensation. In the terminology of
\citeauthor{CoxDonnelly} [(\citeyear{CoxDonnelly}), pages 18--19], these sections constitute units
of analysis, and individual angles are units of interpretation. We
expect interference for angles near neighboring sections. Interference
should be substantial for a large difference in neighboring
compensations, and negligible otherwise.

This reasoning led to the following restricted Latin square design to
study interference. We apply compensations to four cylinders of radius
$0.5, 1, 2$, and $3$ inches, with each cylinder divided into $16$
equal-sized sections of $\pi/8$ radians. One unit of compensation is
$0.004, 0.008, 0.016$, and $0.03$ inch for each respective cylinder,
and there are only four possible levels of compensation, $-1, 0, +1$,
and $+2$ units. Two blocking factors are considered. The first is the
quadrant and the second is the ``symmetry group'' consisting of $\pi/8$-radian sections that are reflections about the coordinate axes from
each other. Symmetric sections form a meaningful block: if compensation
$x$ is applied to all units, then we have from (\ref{eqmodelcompcylinder}) that for $0 \leq\theta\leq\pi/2$,
\begin{eqnarray*}
\mathbb{E} \bigl\{ \Delta r(\theta, r_0, x\mathbf{1}) \vert \alpha,
\beta, a, b, x_0, \sigma \bigr\} &=& \mathbb{E} \bigl\{ \Delta r(\pi-
\theta, r_0, x\mathbf{1}) \vert \alpha, \beta, a, b,
x_0, \sigma \bigr\}
\\
&= & \mathbb{E} \bigl\{ \Delta r(\pi+ \theta, r_0, x\mathbf{1}) \vert
\alpha, \beta, a, b, x_0, \sigma \bigr\}
\\
&=&  \mathbb{E} \bigl\{ \Delta r(2\pi- \theta, r_0, x\mathbf{1}) \vert \alpha, \beta, a, b, x_0, \sigma \bigr\},
\end{eqnarray*}
suggesting a need to control for this symmetry in the experiment. Thus,
for each product, we conceive of the $16$ sections as a $4 \times4$
table, with symmetry groups forming the column blocking factor and
quadrants the row blocking factor. Based on prior concerns about the
possible severity of interference and resulting scope of inference from
our model (\ref{eqcomptaylor}), the set of possible designs was
restricted to Latin squares (each compensation level occurs only once
in any quadrant and symmetry group), where the absolute difference in
assigned treatments between two neighboring sections does not exceed
two levels of compensation. Each product was randomly assigned one
design from this set, with no further restriction that all the products
have the same design.

\begin{figure}

\includegraphics{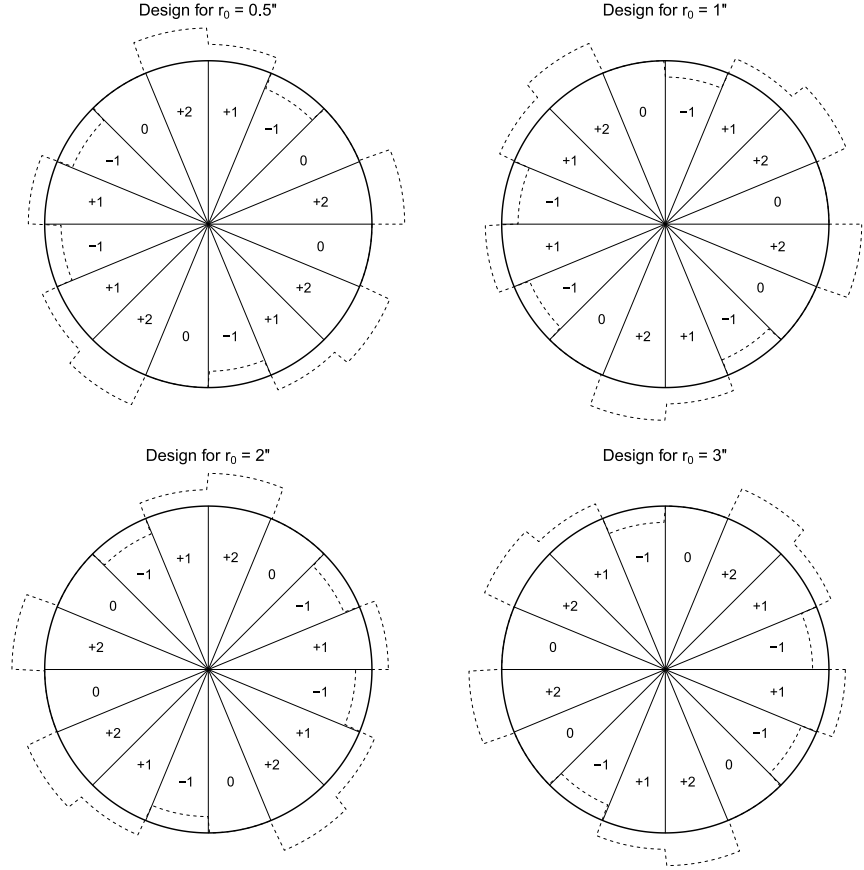}

\caption{Experimental designs. Dashed lines represent assigned compensations.}
\label{design}
\end{figure}

\begin{figure}

\includegraphics{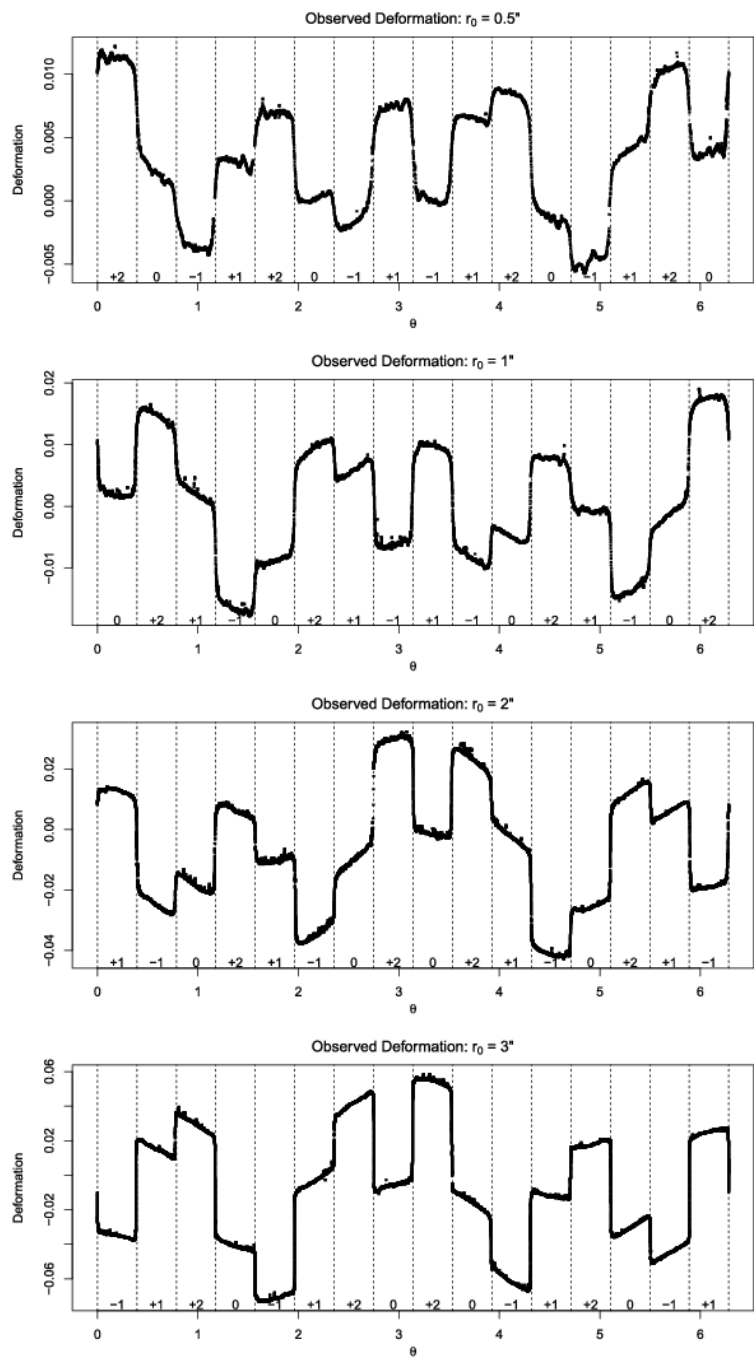}

\caption{Observed deformations in the experiment. Dashed lines
represent sections, and numbers at the bottom of each represent
assigned compensations.}
\label{experimentaldata}
\end{figure}

Our restricted Latin square design forms a discretized compensation
plan that blocks on two factors suggested by the previous deformation
model, and remains model-robust to a certain extent. The chosen
experimental designs are in Figure~\ref{design}, and observed
deformations for the manufactured products are in Figure~\ref{experimentaldata}. There are $N_{0.5} = 6159, N_1 = 6022, N_2 =
6206$, and $N_3 = 6056$ equally spaced angles considered for the four cylinders.

\subsection{Assessing the structure of interference}
\label{secassessinginterference}

Our first task is to assess which units have negligible interference in
the experiment. To do so, we use the suggestions of \citet{sobel}
and \citet{rosenbaum}, who describe when interest exists in
comparing a treatment assignment $\mathbf{x}$ to a baseline.

We have in Section~\ref{secnocompensationfit} data on cylinders that
receive no compensation (denoted by $\mathbf{D}_{n}$) and a model (\ref
{eqnocompmodel}) that provides a good fit. Furthermore, we have a
hypothesized model (\ref{eqmodelcompcylinder}) for compensation
effects when interference is negligible, which is a function of
parameters in (\ref{eqnocompmodel}). If the manufacturing process is
in control, posterior inferences based on $\mathbf{D}_n$ then yield, by
(\ref{eqmodelcompcylinder}), predictions for the experiment. In the
absence of any other information, units in the experiment with observed
deformations deviating strongly from their predictions can be argued to
have substantial interference. After all, if $\theta_i$ has negligible
interference under assignment $\mathbf{x} = (x_1, \ldots, x_N)$, then
\[
\Delta r(\theta_i, r_0, \mathbf{x}) = \Delta r\bigl(
\theta_i, r_0, (x_i, \ldots,
x_i)\bigr) = \Delta r(\theta_i, r_0,
x_i\mathbf{1}).
\]
This suggests the following procedure to assess interference:

\begin{longlist}[(3)]
\item[(1)] Calculate the posterior distribution of the parameters
conditional on $\mathbf{D}_{n}$, denoted by
$\pi(\alpha, \beta, a, b, x_0, \sigma\vert \mathbf{D}_{n})$.

\item[(2)] For every angle in the four cylinders, form the posterior
predictive distribution of the potential outcome
corresponding to the observed treatment assignment (Figure~\ref{design}) using model
(\ref{eqmodelcompcylinder}) and $\pi(\alpha, \beta, a, b, x_0,
\sigma\vert \mathbf{D}_{n})$.
%
\item[(3)] Compare the posterior predictive distributions to the observed
deformations in the experiment.
\begin{itemize}
\item If a unit's observed outcome falls within the $99\%$ central
posterior predictive interval and follows
the posterior predictive mean trend, it is deemed to have negligible
interference.

\item Otherwise, we conclude that the unit has substantial interference.
\end{itemize}
\end{longlist}

This procedure is similar to the construction of control charts [\citet{boxetal}]. When an observed outcome lies outside the $99\%$ central
posterior predictive interval, we suspect existence of a special cause.
As the entire product is manufactured simultaneously, we believe that
the only reasonable assignable cause is interference.

We implemented this procedure and observed that approximately 70\%--80\% of units, primarily in the central regions of sections, have
negligible interference (Appendix~\ref{secposteriorpredictivecheck}). This is clearly seen with another
graph that assesses effective treatments, which we proceed to describe.

\begin{figure}

\includegraphics{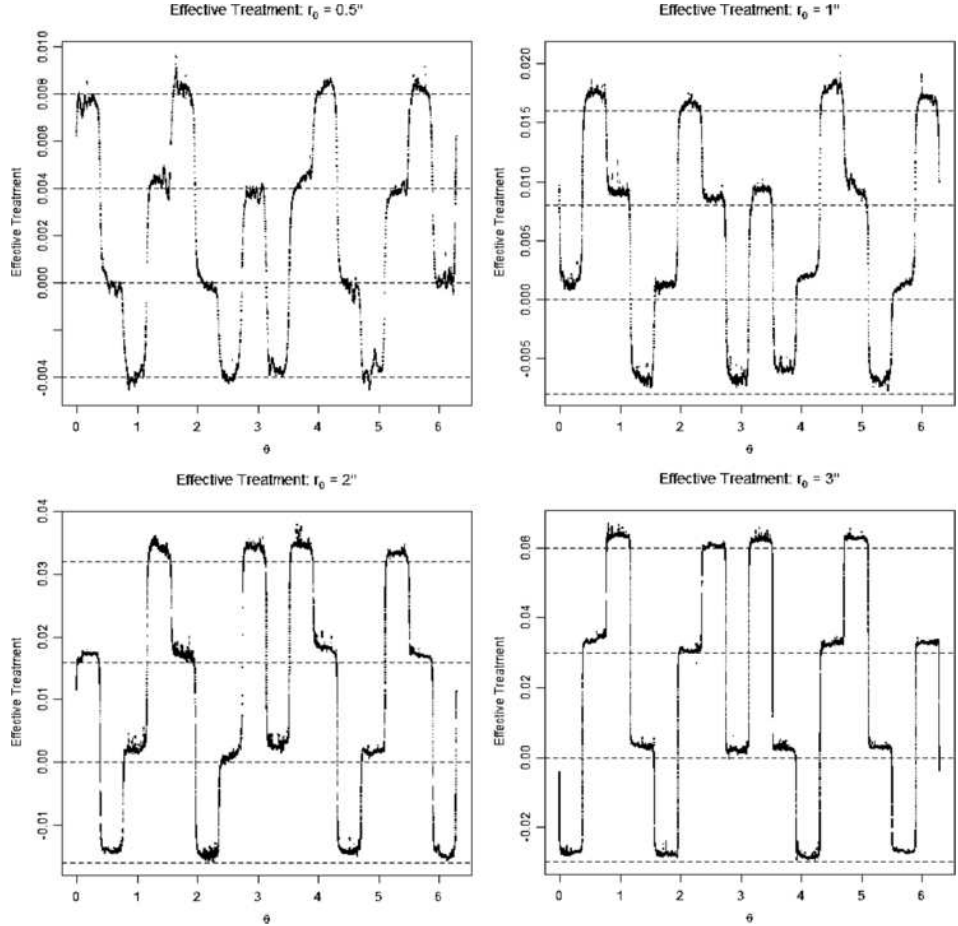}

\caption{Gauging effective treatment $g_i(\mathbf{x})$ using (\protect\ref{eqtreatmentinterference}). Four horizontal lines in each subfigure
denote the possible compensations, and dots denote estimates of
treatments that units effectively received in the experiment.}
\label{posteriorpredictivetreatment}
\end{figure}

Taking expectations in (\ref{eqmodelcompcylinder}), the treatment
effectively received by $\theta_i$ is
\begin{eqnarray}
\label{eqtreatmentinterference}
&& \bigl( \mathbb{E}  \bigl\{ \Delta r(\theta_i, r_0, \mathbf{x}) \vert
\alpha, \beta, a, b, x_0, \sigma \bigr\} -
x_0 \nonumber\\
&& \qquad {}- \alpha(r_0 + x_0)^a - \beta(r_0 + x_0)^b\cos(2 \theta
_i) \bigr)\\
&& \qquad  \hspace{2pt}/ \bigl( 1 + a \alpha(r_0 + x_0)^{a-1} + b \beta(r_0 + x_0)^{b-1}\cos(2 \theta_i)\bigr).\nonumber
\end{eqnarray}
We gauge $g_i(\mathbf{x})$ by plugging observed data from the
experiment and posterior draws of the parameters based on $\mathbf
{D}_{n}$ into (\ref{eqtreatmentinterference}). These discrepancy
measure [\citet{Meng}] calculations, summarized in Figure~\ref{posteriorpredictivetreatment}, again suggest that central angles in
each section have negligible interference: estimates of their effective
treatments correspond to their assigned treatments. There is a slight
discrepancy between assigned treatments and inferred effective
treatments for some central angles, but this is likely due to different
parameter values for the two data sets. Of more importance is the
observation that the effective treatment of a boundary angle $\theta_i$
is a weighted average of the treatment assigned to its section,
$x_{i,M}$, and its nearest neighboring section, $x_{i,\mathit{NM}}$, with the
weights a function of the distances (in radians) between $\theta_i$ and
the midpoint angle of its section, $\theta_{i,M}$, and the midpoint
angle of its nearest neighboring section, $\theta_{i,\mathit{NM}}$. All these
observations correspond to the intuition that interference should be
substantial near section boundaries.

\subsection{A simple interference model}
\label{secsimplemodelinterference}

We first alter (\ref{eqmodelcompcylinder}) to
%
\begin{eqnarray}
\label{eqfullmodelcylinder}
 && \Delta r(\theta_i, r_0, \mathbf{x})\nonumber\\
&& \qquad =
x_0 + \alpha(r_0+x_0)^a +
\beta (r_0+x_0)^b \cos(2
\theta_i)
\\
&&\qquad\quad {}+ \bigl\{ 1 + a\alpha(r_0+x_0)^{a-1}
+ b\beta(r_0+x_0)^{b-1}\cos (2
\theta_i) \bigr\} g_i(\mathbf{x}) +
\varepsilon_{i},\nonumber
\end{eqnarray}
where
%
\begin{eqnarray}
\label{eqweightedtreatment}
g_i(\mathbf{x}) &=& \bigl\{ 1 + \exp  \bigl( -
\lambda_{r_0} | \theta _i - \theta_{i,\mathit{NM}}| +
\lambda_{r_0} |\theta_i - \theta_{i,M}| \bigr) \bigr\}^{-1} x_{i,M}
\nonumber
\\[-8pt]
\\[-8pt]
\nonumber
&&{}+ \bigl\{ 1 + \exp \bigl( \lambda_{r_0} |
\theta_i - \theta_{i,\mathit{NM}}| - \lambda_{r_0} |
\theta_i - \theta_{i,M}| \bigr) \bigr\}^{-1}
x_{i,\mathit{NM}},
\end{eqnarray}
with $\theta_{i,M}, \theta_{i,\mathit{NM}}$ denoting midpoint angles for the $\pi
/8$-radian sections containing and neighboring nearest to $\theta_i$,
respectively, and $x_{i,M}, x_{i,\mathit{NM}}$ compensations assigned to these
sections. Effective treatment $g_i(\mathbf{x})$ is a weighted average
of the unit's assigned treatment $x_i = x_{i,M}$ and the treatment
$x_{i,\mathit{NM}}$ assigned to its nearest neighboring section. Although the
form of the weights is chosen for computational convenience, we
recognize that (\ref{eqweightedtreatment}) belongs to a class of
models agreeing with prior subject-matter knowledge that interference
may be negligible if the implemented compensation plan is sufficiently
``continuous,'' in the sense that the theoretical compensation plan is
a continuous function of $\theta$ and the tolerance of the 3D printer
is sufficiently fine so that discretization of compensation is
negligible (Appendix~\ref{secnote}).

We fit the model in (\ref{eqfullmodelcylinder}) and (\ref
{eqweightedtreatment}), having $10$ total parameters, to the
experiment data. The prior specification remains the same, with $\log (\lambda_{r_0}) \sim\mathrm{N}(0,4^2)$ independently {a
priori} for $r_0 = 0.5, 1, 2$, and $3$ inches. A HMC algorithm was used
to obtain $1000$ draws from the joint posterior distribution after a
burn-in of $500$, and these are summarized in Table~\ref{posteriorpredictivetableexperimental}.

\begin{table}
\centering
\caption{Summary of posterior draws for simple interference model}
\label{posteriorpredictivetableexperimental}
\begin{tabular*}{\tablewidth}{@{\extracolsep{\fill}}ld{2.9}d{1.9}d{2.9}d{3.15}c@{}} \hline
     & \multicolumn{1}{c}{\textbf{Mean}} & \multicolumn{1}{c}{\textbf{SD}} & \multicolumn{1}{c}{\textbf{Median}} & \multicolumn{1}{c}{$\bolds{95\%}$ \textbf{credible}
     \textbf{interval}} & \multicolumn{1}{c@{}}{\textbf{ESS}} \\
     \hline
$\alpha$              & -1.06 \times 10^{-2}              & 1.53 \times 10^{-4}      & -1.06 \times 10^{-2}     & (-1.09,-1.03) \times 10^{-2}         & 8078    \\
$\beta$               & 5.79 \times 10^{-3}               & 3.69 \times 10^{-5}      & 5.79 \times 10^{-3}      & (5.72, 5.86) \times 10^{-3}          & 8237    \\
$a$                   & 9.5 \times 10^{-1}                & 9.46 \times 10^{-3}      & 9.5 \times 10^{-1}       & (9.31, 9.69) \times 10^{-1}          & 8150    \\
$b$                   & 1.12                              & 6.64 \times 10^{-3}      & 1.12                     & (1.0, 1.13)                          & 8504    \\
$x_0$                 & 7.1 \times 10^{-3}                & 1.43 \times 10^{-4}      & 7.1 \times 10^{-3}       & (6.82, 7.39) \times 10^{-3}          & 8404    \\
$\sigma$              & 3.14 \times 10^{-3}               & 1.36 \times 10^{-5}      & 3.14 \times 10^{-3}      & (3.11, 3.17) \times 10^{-3}          & 8924    \\
$\lambda_{0.5}$       & 32.66                             & 2.05                     & 32.62                    & (28.69, 36.76)                       & 8686    \\
$\lambda_{1}$         & 48.24                             & 2                        & 48.12                    & (44.5, 52.6)                         & 8666    \\
$\lambda_{2}$         & 76.83                             & 1.78                     & 76.78                    & (73.42, 80.44)                       & 8770    \\
$\lambda_{3}$                                                                        & 86.08
& 0.83
& 86.06                               & (84.49, 87.68)
& 8385
\\
\hline
\end{tabular*}
\end{table}

\begin{figure}

\includegraphics{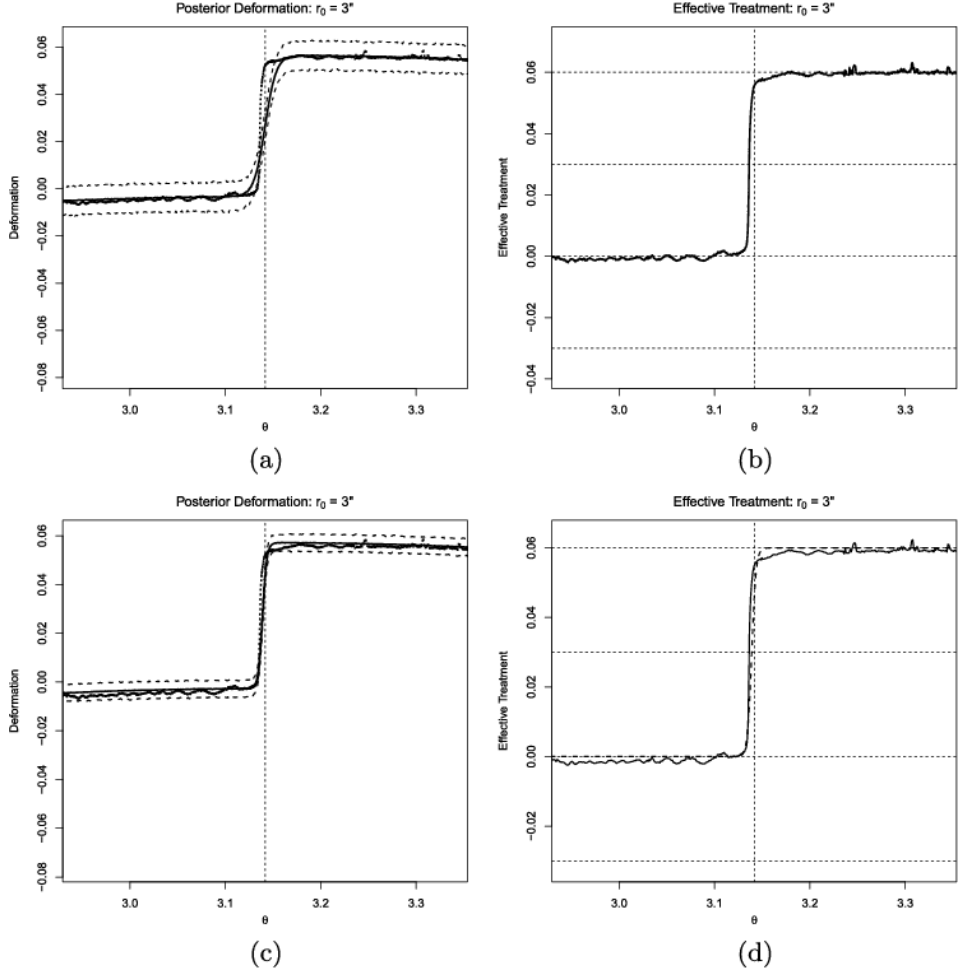}

\caption{\textup{(a)} An example of the type of erroneous predictions
made by model (\protect\ref{eqfullmodelcylinder}),
(\protect\ref{eqweightedtreatment}) for the $3$ inch cylinder. The vertical
line is drawn at $\theta= \pi$, marking the boundary between two
sections. Units to the left of this line were given $0$ compensation,
and units to the right were given $+2$ compensation. The posterior mean
trend is represented by the solid line, and posterior quantiles are
represented by dashed lines. Observed data are denoted by dots.
\textup{(b)} Corresponding inferred effective treatment for $15\pi/16
\leq\theta\leq17\pi/16$.
\textup{(c)} Refined posterior predictions for $r_0 = 3$ inches, $15\pi
/16 \leq\theta\leq17\pi/16$.
\textup{(d)} Comparing inferred effective treatments (solid line) with
refined effective treatment model (dashed line) for
the $3$ inch cylinder.}
\label{posteriorpredictive3error}
\end{figure}

This model provides a good fit for the $0.5$ and $1$ inch cylinders,
but not the others. As an example, in Figure~\ref{posteriorpredictive3error}(a) the posterior mean trend does not
correctly capture the observed transition across sections for the $3$
inch cylinder. The problem appears to reside in (\ref{eqweightedtreatment}). This specification implies that effective
treatments of units $\theta_i = k\pi/8$ for $k \in\mathbb{Z}_{>0}$ are
equal-weighted averages of compensations applied to units $k\pi/8 \pm
\pi/16$. To assess the validity of this implication, we use the
posterior distribution of the parameters to calculate, for each $\theta
_i$, the inferred effective treatment in (\ref{eqtreatmentinterference}).
An example of these calculations, Figure~\ref{posteriorpredictive3error}(b), shows that the inferred
effective treatment for $\theta_i = \pi$ is nearly $0.06$ inch, the
compensation applied to the right-side section. Thus, specification
(\ref{eqweightedtreatment}) is invalidated by the experiment.

Another posterior predictive check helps clarify the problem. From (\ref{eqweightedtreatment}),
\[
g_i(\mathbf{x}) = w_i x_{i,M} + (1 -
w_i) x_{i,\mathit{NM}},
\]
and so
%
\begin{equation}\label{eqinferredweightfunction}
w_i = \frac{g_i(\mathbf{x}) - x_{i,\mathit{NM}}}{x_{i,M} - x_{i,\mathit{NM}}},
\end{equation}
which is well defined because $x_{i,M} \neq x_{i,\mathit{NM}}$ in this
experiment. Plugging in the inferred effective treatments, calculated
from (\ref{eqtreatmentinterference}), into (\ref{eqinferredweightfunction}), we then diagnose how to modify (\ref{eqweightedtreatment}) to better model
interference in the experiment.

\begin{figure}

\includegraphics{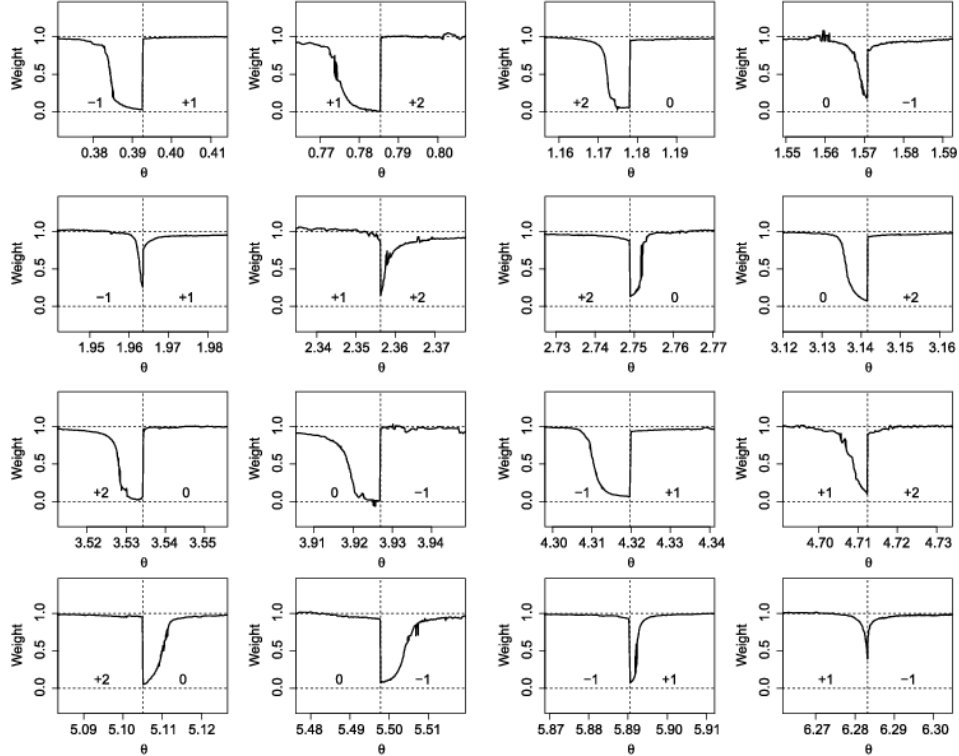}

\caption{Inferring weights $w_i$ in the interference model for the $r_0
= 3$ inch cylinder, using effective treatments calculated from equation
(\protect\ref{eqtreatmentinterference}), based on the posterior distribution
of parameters from Section~\protect\ref{secsimplemodelinterference} and
equation (\protect\ref{eqinferredweightfunction}). Vertical lines represent
$\theta= k\pi/8$ for $k = 1, \ldots, 16$, and numbers at the bottom of
each subfigure represent assigned compensations.}
\label{radius3weight}
\end{figure}

This calculation was made for all cylinders, and the results for $r_0 =
3$ inches are summarized in Figure~\ref{radius3weight} as an example.
Rows in this figure show the weights for each quadrant, and we focus on
their behavior in neighborhoods of integral multiples of $\pi/8$.
Neither the decay in the weights [represented by $\lambda_{r_0}$ in
(\ref{eqweightedtreatment})] nor the weight for integral multiples of
$\pi/8$ remain constant across sections. In fact, these figures suggest
that $\lambda_{r_0}$ is a function of $\theta_{i,M}, \theta_{i,\mathit{NM}}$,
and that a location term is required. They also demonstrate a possible,
subtle quadrant effect and, as our experiment blocks on this factor, we
are better able to use these posterior predictive checks to refine our
simple interference model and capture this unexpected deformation pattern.

\subsection{A refined interference model}
\label{secrefinedmodelinterference}

Our refined effective treatment model is of the same form as (\ref
{eqweightedtreatment}), with $\lambda_{r_0}$ replaced by $\lambda
_{r_0}(\theta_{i,M}, \theta_{i,\mathit{NM}})$, and $|\theta_i - \theta_{i,M}|,
|\theta_i - \theta_{i,\mathit{NM}}|$ replaced by $|\theta_i - \theta_{i,M} -
\delta_{r_0}(\theta_{i,M}, \theta_{i,\mathit{NM}})|, |\theta_i - \theta_{i,\mathit{NM}} -
\delta_{r_0}(\theta_{i,M}, \theta_{i,\mathit{NM}})|$, respectively. Here, $\delta
_{r_0}(\theta_{i,M}, \theta_{i,\mathit{NM}})$ represent location shifts across
sections suggested by the previous posterior predictive checks.

Our specific model is
%
\begin{eqnarray}
\label{eqdeltaFourierexpansion}
\delta_{r_0}(\theta_{i,M},
\theta_{i,\mathit{NM}}) & =&  \delta_{r_0,0} + \sum
_{k=1}^3 \bigl\{ \delta_{r_0,k}^c
\cos(k\theta_{i,B}) + \delta_{r_0,k}^s
\sin(k\theta_{i,B}) \bigr\},
\\
\label{eqlambdarefined}
\lambda_{r_0}(\theta_{i,M},
\theta_{i,\mathit{NM}}) &=& \mathbb{I}\bigl(|x_{i,M} - x_{i,\mathit{NM}}| = 1\bigr)
\lambda_{r_0,1}
\nonumber
\\[-8pt]
\\[-8pt]
&&{}+ \mathbb{I}\bigl(|x_{i,M} - x_{i,\mathit{NM}}| = 2\bigr)
\lambda_{r_0,2},\nonumber
\end{eqnarray}
where $\theta_{i,B} = (\theta_{i,M} + \theta_{i,\mathit{NM}})/2$ and $|x_{i,M} -
x_{i,\mathit{NM}}|$ is measured in absolute units of compensation. From Figure~\ref{radius3weight} and the fact that
\[
\delta_{r_0}(\theta_{i,M}, \theta_{i,\mathit{NM}}) =
\delta_{r_0}(\theta_{i,M} + 2\pi, \theta_{i,\mathit{NM}} + 2\pi),
\]
location shifts should be modeled using harmonic functions.

This model provides a better fit. Comparing Figure~\ref{posteriorpredictive3error}(c), which displays posterior predictions
from the refined model (based on one chain of posterior draws using a
standard random walk Metropolis algorithm), with the previous model's
predictions in Figure~\ref{posteriorpredictive3error}(a), we
immediately see that the refined model better captures the posterior
predictive mean trend. Similar improvements exist for the other
sections and cylinders. We also compare the original inferred effective
treatments obtained from (\ref{eqtreatmentinterference}) with the
refined model in Figure~\ref{posteriorpredictive3error}(d) and again
observe that the new model better captures interference.


\subsection{Summary of the experimental design and analysis}
\label{secdiscussion}

Three key ingredients relating to the data, model, and experimental
design have made our series of analyses possible, and are relevant and
useful across a wide variety of disciplines. First is the availability
of benchmark data, for example, every unit on the cylinder receiving
zero compensation. Second is the potential outcomes model~(\ref{eqmodelcompcylinder}) for compensation effects when there is no
interference, defined in terms of a fixed number of parameters that do
not depend on the compensation plan $\mathbf{x}$. These two enable
calculation of the posterior predictive distribution of potential
outcomes under the assumption of negligible interference. The final
ingredient is the explicit distinction between units of analysis and
units of interpretation in our design, which provides the means to
assess and model interference in the experiment. Comparing observed
outcomes from the experiment to posterior predictions allows one to
infer the structure of interference, which can be validated by further
experimentation.

These considerations suggest that our methodology can be generalized
and applied to other experimental situations with units residing on
connected surfaces. In general, when experimenting with units on a
connected surface, a principled and step-by-step analysis using the
three ingredients above, as illustrated in this paper, can ultimately
shed more light on the substantive question of interest.

\section{Conclusion: Ignoring interference inhibits improvements}
\label{sec4}

To manufacture 3D printed products satisfying dimensional accuracy
demands, it is important to address the problem of interference in a
principled manner.
\citet{huangzhangsabbaghidasgupta} recognized that continuous
compensation plans implemented on printers with a sufficiently fine
tolerance can effectively control a product's printed dimensions
without inducing additional complications through interference. Their
models for product deformation motivated our experiment that introduces
interference through the application of a discretized compensation plan
to the boundary of a cylinder. Combining this experiment's data with
inferences based on data for which every unit received no compensation
led to an assessment of interference in terms of how units' effective
treatments differed from that physically assigned. Further analyses
effectively modeled interference in the experiment.

It is important to note that the refined interference model's location
and scale terms (\ref{eqdeltaFourierexpansion}), (\ref
{eqlambdarefined}) are a function of the compensation plan. For
example, reflecting the assigned compensations across the y axis would
accordingly change the location shifts. The implication of this and all
our previous observations for manufacturing is that severely
discretized compensation plans introduce interference, and, if this
fact is ignored, then quality control of 3D printed products will be
hindered, especially for geometrically complex products relevant in
real-life manufacturing.

Many research challenges and opportunities for both statistics and
additive manufacturing remain to be addressed. Perhaps the most
important is experimental design in the presence of interference. For
example, when focus is on the construction of specific classes of
products (e.g., complicated gear structures), optimum designs can lead
to precise estimates of model parameters, hence improved compensation
plans and control of deformation. An important and subtle statistical
issue that then arises is how the structure of interference changes as
a function of the compensation plan derived from the experimental
design. Instead of being a weighted average of the treatment applied to
its section and nearest neighboring section, the derived compensation
plan may cause a unit's effective treatment to be a weighted average of
treatments applied to other sections as well, with weights depending on
the absolute difference in applied compensations. Knowledge of the
relationship between compensation plans derived from specific
experimental designs and interference is necessary to improve quality
control in general, and therefore is an important issue to address for
3D printing.

\begin{appendix}

\section{Correlation in \texorpdfstring{$\varepsilon$}{varepsilon}}
\label{seccorrelation}

In all our analyses, we assumed the $\varepsilon_i$ were independent. As
pointed out by a referee, when units reside on a constrained boundary,
independence of error terms is generally unrealistic. However, we
believe that our specific context helps justify this simplifying
assumption for several reasons.

First, the major objective driving our work on 3D printing is
compensation for product deformation. To derive compensation plans, it
is important to accurately specify the mean trend in deformation.
Although incorporating correlation may change parameter estimates that
govern the mean trend, we do not believe that modeling the correlation
in errors will substantially help us compensate for printed product
deformations. This is something we intend to address further in our
future work.

\begin{figure}

\includegraphics{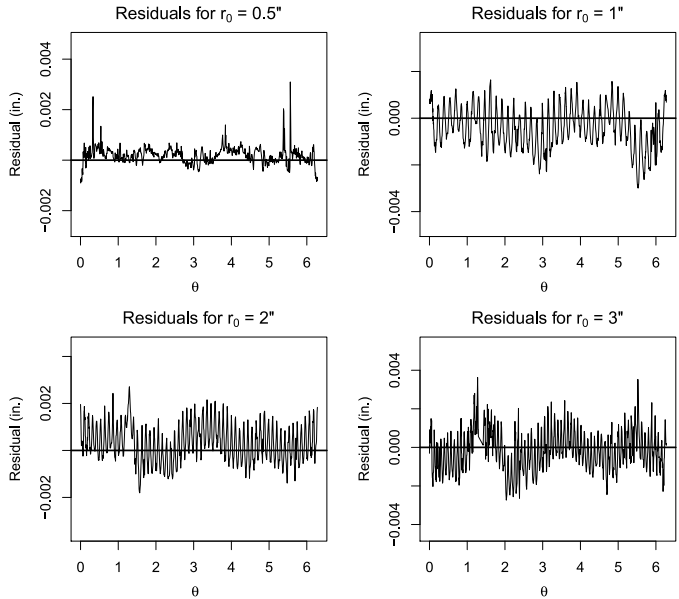}

\caption{Residuals for the model fit in Section~\protect\ref{secnocompensationfit}. Here, the residual is defined as the
difference between the observed deformation and the posterior mean of
deformation for each angle $\theta_i$.}\vspace*{-5pt}
\label{figresiduals}
\end{figure}

Second, there is a factor that may further confound the potential
benefits of including correlated errors in our model: the resolution of
the CAD model. To illustrate, consider the model fit in Section~\ref{secnocompensationfit}.
We display the residual plots in Figure~\ref{figresiduals}. All residuals are (in absolute value) less than $1\%$
of the nominal radius for $r_0 = 0.5$ inch and at most approximately
$0.1\%$ of the nominal radius for $r_0 = 1, 2, 3$ inches, supporting
our claim that we have accurately modeled the mean trend in deformation
for these products. However, we note that for $r_0 = 1, 2, 3$ inches,
there is substantial negative correlation in residuals between adjacent
units, with the residuals following a high-frequency harmonic trend.
There is a simple explanation for this phenomenon. Our first
manufactured products were $r_0 = 1, 2, 3$ inches, and the CAD models
for these products had low resolution. Low resolution in the CAD model
yields the high-frequency pattern in the residual plots. The next
product we constructed was $r_0 = 0.5$ inch, and its CAD model had
higher resolution than that previously used, which helped to remove
this high-frequency pattern. Minor trends appear to exist in this
particular plot, and an ACF plot formally reveals significant
autocorrelations. Accordingly, we observe that the correlation in
residuals is a function of the resolution of the initial CAD model. In
consideration of our current data and our primary objective to
accurately capture the mean trend in deformation, we use independent
$\varepsilon_i$ throughout. We intend to pursue this issue further in our
future work, for example, in the direction of \citet
{colosimosemeraropacella}.

Furthermore, as pointed out by the Associate Editor, correlations in
residuals for more complicated products may be accounted for by
modeling the interference between units, which is precisely the focus
of this manuscript.

\section{MCMC convergence diagnostics}
\label{secdiagnostics}

Convergence of our MCMC algorithms was gauged by analysis of ACF and
trace plots, and effective sample size (ESS) and \citeauthor{GelmanRubin} [(\citeyear{GelmanRubin}), GR] statistics, which were calculated using $10$
independent chains of $1000$ draws after a burn-in of $500$. In
Sections~\ref{secnocompensationfit} and \ref{secsimplemodelinterference}, the ESS were all above $8000$ (the
maximum is $10\mbox{,}000$), and the GR statistics were all $1$.

\section{Assessing interference}
\label{secposteriorpredictivecheck}

\begin{figure}

\includegraphics{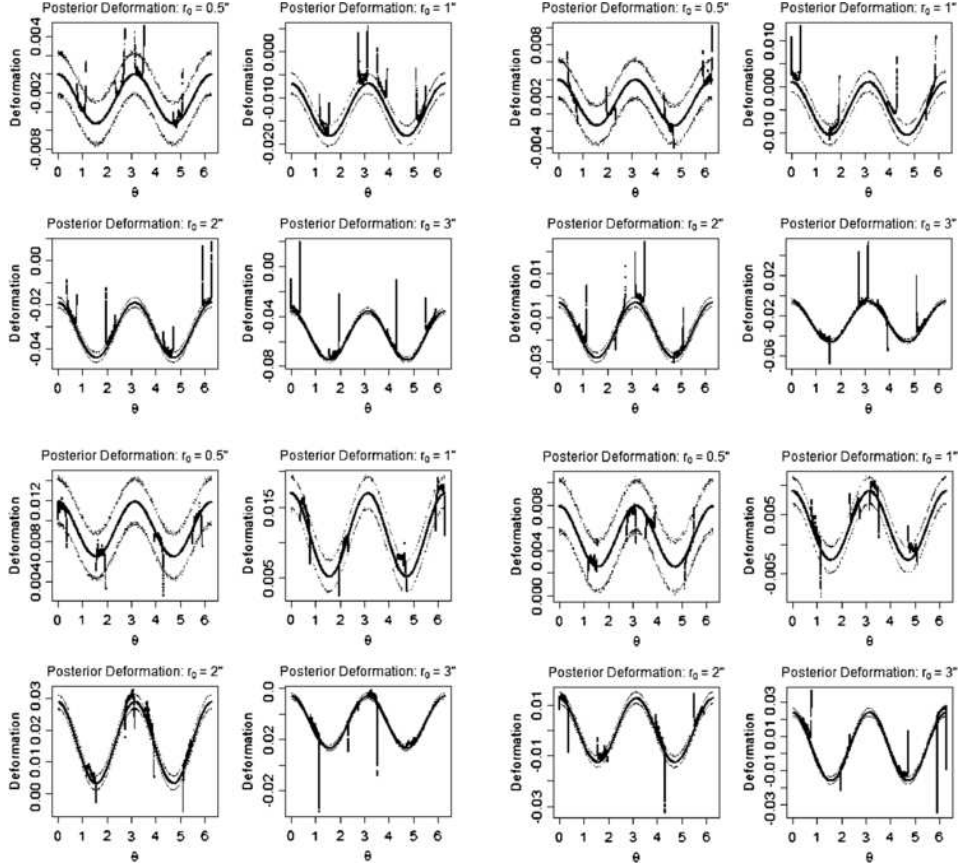}

\caption{Assessing interference in the experiment based on posterior
inferences drawn from the no-compensation data.
Clockwise from top left: predictions for units that received $-1, 0,
+1$, and $+$2 compensation.}\vspace*{-6pt}
\label{posteriorpredictiveexperiment}
\end{figure}

The results of the first procedure described in Section~\ref{secassessinginterference} are displayed in Figure~\ref{posteriorpredictiveexperiment}: bold lines represent posterior
means, dashed lines quantiles forming the  99\%  central posterior\vadjust{\goodbreak}
intervals, and dots the observed outcomes in the experiment, with
separate figures for each nominal radius and compensation. For example,
the graph in the first row and column of Figure~\ref{posteriorpredictiveexperiment} contains the observed data for angles
in the $0.5$ inch radius cylinder that received $-$1 compensation. This
figure also contains the posterior predictive mean and  99\%
intervals for all angles under the assumption that $-$1 compensation
was applied uniformly to the cylinder. Although only four sections of
the cylinder received this compensation in the experiment, forming this
distribution makes the posterior predictive mean trend transparent, and
so helps identify when a unit's observed outcome deviates strongly from
its prediction.

\section{Note on a class of interference models}
\label{secnote}

Compensation is applied in practice by discretizing the plan at a
finite number of points, according to some tolerance specified\vadjust{\goodbreak} by the
size (in radians) for each section or, alternatively, the maximum value
of $|\theta_{i,M} - \theta_{i,\mathit{NM}}|$.

Suppose compensation plan $x(\theta)$ is a continuous function of
$\theta$, and define
\[
w_i = \frac{h(|\theta_i - \theta_{i,M} |)}{
h(|\theta_i - \theta_{i,M} |) + h(|\theta_i - \theta_{i,\mathit{NM}}|)},
\]
with $h\dvtx \mathbb{R} \rightarrow\mathbb{R}_{>0}$ a monotonically
decreasing continuous function, and
\[
g_i(\mathbf{x}) = w_i x_{i,M} + (1 -
w_i) x_{i, \mathit{NM}}.
\]
Then for the cylinders considered in our experiment, $g_i(\mathbf{x})
\rightarrow x_i$ as $|\theta_{i,M} - \theta_{i, \mathit{NM}}| \rightarrow0$.
This is because $|x_{i,M} - x_{i,\mathit{NM}}| \rightarrow0$ as $|\theta_{i,M}
- \theta_{i,\mathit{NM}}| \rightarrow0$, and
\[
0 \leq |\theta_i - \theta_{i,\mathit{NM}} | - |\theta_i
- \theta_{i,M} | \leq |\theta_{i,M} - \theta_{i,\mathit{NM}}|.
\]
\end{appendix}

\section*{Acknowledgements}
We are grateful to Xiao-Li Meng, Joseph Blitzstein, David Watson,
Matthew Plumlee, the Editor, Associate Editor, and a referee for their
valuable comments, which improved this paper.


%



\printaddresses
\end{document}